\documentclass[epj]{svjour}

\newcommand{\Vista}{{\sc Vista}}
\newcommand{\Sleuth}{{\sc Sleuth}}
\newcommand{\sumPt}{{\ensuremath{\sum{p_T}}}}

\newcommand{\vecpTslash}{{\ensuremath{\vec{p}\!\!\!/_T}}}

\newcommand{\scriptP}{\ensuremath{{\cal P}}}
\newcommand{\tildeScriptP}{\ensuremath{ {\cal \tilde{P}} }}

\usepackage{graphicx}
\usepackage{fancyhdr}
\usepackage{amsmath}
\usepackage{amssymb}
\def\makeheadbox{{
 }}
\def\mytitle{My title} 
\def\myauthors{My name}  
\def\mytype{My type of session}
\def\mysession{My session}

\def\mytitle{Sleuth at CDF} 
\def\myauthors{Georgios Choudalakis}    
\def\mytype{Contributed Talk}    
\def\mysession{Alternatives}

\begin{document}
\title{Sleuth at CDF}
\subtitle{A quasi-model-independent search for new electroweak scale physics}
\author{Georgios Choudalakis\thanks{http://www.mit.edu/$\sim$gchouda/, 
gchouda@mit.edu}%
}                     
%
%
\institute{Massachusetts Institute of Technology, for the CDF collaboration}
%
\date{}
\abstract{
These proceedings describe \Sleuth, a quasi-model-independent search strategy targeting new electroweak scale physics, and its application to 927~pb$^{-1}$ of CDF II data.  Exclusive final states are analyzed for an excess of data beyond the Standard Model prediction at large summed scalar transverse momentum ($\sumPt$).  This analysis of high-$\sumPt$ data represents one of the most encompassing searches so far conducted for new physics at the energy frontier.
\PACS{
      {12.60.-i}{Models beyond the standard model}  
     } 
} 
\maketitle
\section{Introduction}
\label{intro}

Searching a large subset of collision events recorded at the energy frontier for evidence of new physics requires understanding the Standard Model prediction and detector response.  To facilitate understanding, debris from collision events are reduced to 4-vectors of reconstructed physics objects:  electrons ($e$), muons ($\mu$), taus ($\tau$), photons ($\gamma$), jets ($j$), and jets tagged as containing a bottom quark ($b$).  Existing event generators are used to construct the Standard Model prediction, and the response of the CDF detector to these events is estimated using a {\sc{geant}}-based simulation of the detector.  A correction model is developed to correct possible inadequacies in the Standard Model prediction and modeling of the detector response.  These procedures form the basis of  \Vista, described in a companion proceedings in this conference~\cite{vista}.   Having obtained a global Standard Model background estimate, \Sleuth\ is used to target new physics likely to appear as an excess of data beyond Standard Model prediction at large summed scalar transverse momentum.  Similar search strategies have been developed previously at D\O~\cite{sleuthPRL,sleuthPRD1,sleuthPRD2} and H1~\cite{H1GeneralSearch}.

\section{The Method}
\label{sec:method}

\Sleuth\ is slightly less model-independent search strategy than \Vista, as it invokes three additional assumptions.  The three additional assumptions motivating \Sleuth\ are that new physics will likely appear
\begin{enumerate}
\item as an excess of data over the Standard Model expectation, 
\item at large \sumPt\ with respect to the Standard Model background, and 
\item predominantly in one \Sleuth\ final state. 
\end{enumerate}
\Sleuth\ is expected to be sensitive to any particular incarnation of new physics to the extent that the above criteria are satisfied.

Following these assumptions, \Sleuth\ searches for statistically significant excesses of data in the high-\sumPt\ tails of all high-$p_T$ final states with non-zero Standard Model expectation, or where data are observed.  Final states are defined so as to concentrate more of the signal of possible new physics.  Final states equivalent under a global switch of the two lesser generations (such as $e^+e^-$ and $\mu^+\mu^-$) are merged (into $\ell^+\ell^-$).  Final states equivalent under global charge conjugation (such as $\ell^+\ell^+$ and $\ell^-\ell^-$) are also merged.  Since new physics is expected to produce light quarks ($j$) and bottom quarks ($b$) in pairs, each event is said to contain $n$ jet pairs if the number of reconstructed jets is $2n$ or $2n+1$, and is said to contain $m$ $b$-jet pairs if the number of reconstructed $b$-jets is $2m$, or $2m-1$ and there is another jet in the event, for integer $m$ and $n$. 

\Sleuth\ examines the variable
\begin{equation}
\label{eq:sumpt}
\sumPt = \left( \sum_{i\in{\rm objects}} {\left|\vec{p}_{Ti}\right|}\right) +\left|{\vec{p}_T}_{\rm uncl}\right| + \left|\vecpTslash\right| ,
\end{equation} 
where the first term is the scalar sum of the transverse momentum of all reconstructed objects, the second term is the transverse momentum of energy visible in the detector but not clustered into reconstructed objects, and the third term is the transverse component of any missing energy in the event.  By definition, these quantities satisfy
\begin{equation}
\sum_{i}{\vec{p}_{Ti}} + \vecpTslash + {\vec{p}_T}_{\rm uncl} = \vec{0},
\end{equation}
a two-component vector equation in the spatial coordinates transverse to the axis of the colliding beams.
In addition to the fact that many new physics scenarios involve new massive resonances decaying into Standard Model objects resulting in events with large \sumPt, the variable \sumPt\ is useful in practice since it is easily defined for any final state, it is fairly insensitive to soft physics, including the details of final state parton showering, and its expected distribution can be easily populated on the high tails with additional Monte Carlo events by running event generators with a (slightly lower) threshold on generated \sumPt.

In each final state, \Sleuth\ considers regions stretching from each data point up to infinity in the distribution of \sumPt.  For each region, the Poisson probability ($p$-value) that the Standard Model background in that region would fluctuate up to or above the number of data events observed in that region is calculated.  The most interesting region in each final state is the one with the smallest $p$-value.  \Sleuth\ next produces ensembles of pseudo data events drawn from the \sumPt\ distribution of the Standard Model background, and for each set of pseudo data events finds the most interesting region and corresponding $p$-value.  The fraction (\scriptP) of this ensemble of pseudo experiments in which the most interesting region corresponds to a $p$-value smaller than that of the most interesting region observed in the actual data is determined.  The value \scriptP, a number ranging between zero (very interesting) and unity (not interesting), quantifies the probability that a random fluctuation of the Standard Model background would create any region more interesting than the most interesting region observed in the data in that particular final state.  In each final state considered individually, \scriptP\ is the relevant measure of the statistical significance of the most interesting region on the tail of the \sumPt\ distribution.  This process is repeated for all \Sleuth\ final states populated by three or more data events.

The most interesting final state is identified as the final state with smallest \scriptP, denoted by $\scriptP_{\rm{min}}$.  An ensemble of pseudo experiments involving all final states is then used to find the fraction (\tildeScriptP) of these pseudo experiments that would produce one or more final states more interesting than the most interesting final state in the data.  The number \tildeScriptP, which ranges between zero (very interesting) and unity (not interesting), quantifies the interest of the final state with the most interesting region, accounting for the trials factor associated with having examined many final states.

\Sleuth's discovery threshold is $\tildeScriptP \lesssim 0.001$.  The trials factor is such that \tildeScriptP\ at this threshold typically corresponds to a region with $p$-value of about $10^{-7}$, motivating the default criterion of $5\sigma$ typically employed in our field.

\section{Sensitivity}
\label{sec:sensitivity}

\begin{figure}
\centering
\includegraphics[height=0.45\textwidth,angle=-90]{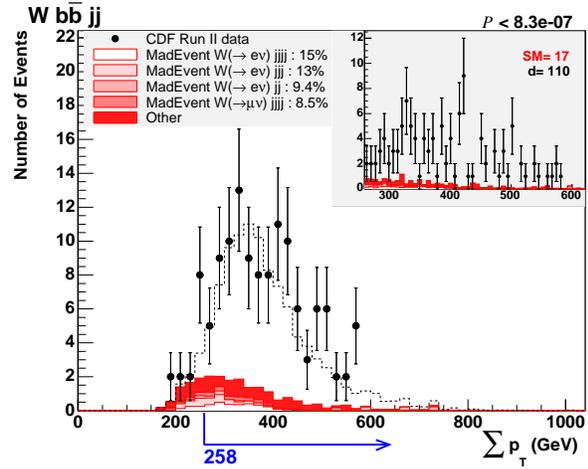} 
\caption{The \Sleuth\ $Wb\bar{b}jj$ final state is dominated by Standard Model $t\bar{t}$ production.  The total Standard Model prediction including $t\bar{t}$ production (dashed line) is in agreement with observed CDF data (black points).  If the top quark were unknown, the Standard Model prediction {\em{sans}} $t\bar{t}$ production (red shaded histogram) significantly underestimates the observed data.  A global fit of the \Vista\ correction factors~\cite{vista} has been unable to coerce the Standard Model prediction without $t\bar{t}$ to match the data in this final state.  The (blue) arrow indicates the most interesting region, identified by \Sleuth.  The interst of this region, considering this final state alone, is found to be $\scriptP < 8.3\times 10^{-7}$ (upper right), corresponding to a value of \tildeScriptP\ easily satisfying \Sleuth's discovery threshold of 0.001.}
\label{fig:topless}
\end{figure}

\begin{figure}
\centering
\includegraphics[width=0.45\textwidth,angle=0]{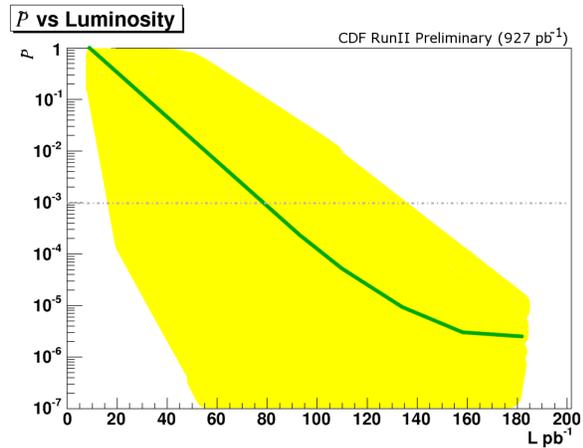} 
\caption{\Sleuth's $\tildeScriptP$ as a function of assumed integrated luminosity, where top quark pair production is removed from the Standard Model background estimate.  The horizontal axis shows integrated luminosity, in units of pb$^{-1}$.  The shaded (yellow) band shows the range of values of $\tildeScriptP$ obtained when the integrated luminosity of the CDF data sample is reduced, with the width of the band resulting from individual CDF top quark candidate events being randomly retained or discarded.  The median $\tildeScriptP$ obtained is shown as the solid (green) line.  The \Sleuth\ discovery threshold $\tildeScriptP \lesssim 0.001$ is shown as the horizontal dashed (gray) line.}
\label{fig:tildeScriptPvsLuminosityTtbar}
\end{figure}

Sensitivity studies are performed to estimate \Sleuth's sensitivity to possible new physics in 927 pb$^{-1}$ of CDF data.   Monte Carlo events corresponding to particular models of new physics, including specific parameter points within the MSSM and several flavors of $Z'$, are partitioned into \Sleuth\ final states according to the objects reconstructed in these events, and are added to pseudo data drawn from the Standard Model prediction.  \Sleuth\ is run ``blind'' on the resulting mix of pseudo data and trace amounts of pseudo signal, not knowing anything about the pseudo signal that has been added.  The amount of pseudo signal is gradually increased until \Sleuth\ identifies a region of interest corresponding to $\tildeScriptP < 0.001$.

The models tested require a cross section of roughly a few picobarns to reach \Sleuth's discovery threshold in 927 pb$^{-1}$ of data.  \Sleuth's sensitivity is found to be comparable to that of targeted searches for models that sufficiently satisfy the three assumptions on which \Sleuth\ is based.  \Sleuth's sensitivity to unspecified new physics is in general significantly greater than any specific targeted search, due to \Sleuth's noticeably broader scope.

Specific Standard Model processes are also used to examine \Sleuth's sensitivity.  If the top quark were unknown, its production at the Tevatron would have been caught using \Sleuth.  Observation of $t\bar{t}$ production in 927~pb$^{-1}$ is shown in Fig.~\ref{fig:topless}. An integrated luminosity of roughly 80~pb$^{-1}$ would be required to reach the \Sleuth\ discovery threshold\footnote{When comparing to the 67~pb$^{-1}$ needed for discovery in CDF Run I~\cite{CDFTopDiscovery}, differences such as the improved CDF II detector should be kept in mind.}, as shown in Fig.~\ref{fig:tildeScriptPvsLuminosityTtbar}.  Standard Model $WW$ production is also easily observed.  \Sleuth\ has more difficulty with single top quark and Higgs boson production, which only partially satisfy \Sleuth's assumption that new physics will appear at large \sumPt\ relative to (other) Standard Model processes.

\begin{figure}
\centering
\includegraphics[height=0.45\textwidth,angle=-90]{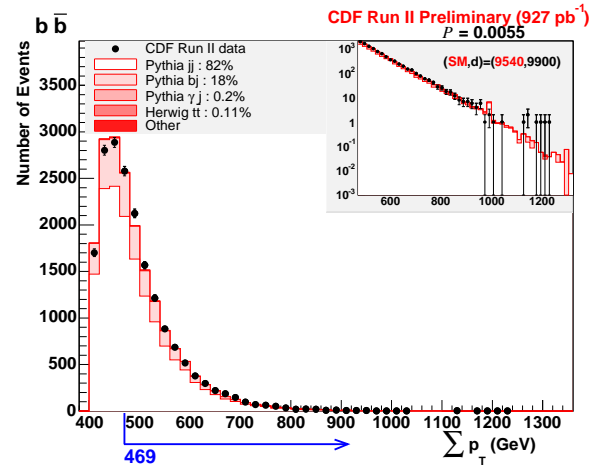} 
\caption{The most interesting \Sleuth\ final state ($b\bar{b}$).  Filled circles (black) show CDF Run II data; the (red) histogram shows the standard model prediction, with $jj$ standing for two non-$b$ generated partons, and $bj$ for at least one generated $b$-quark.  The horizontal axis shows the summed scalar transverse momentum of all objects in each event in the histogram, with the number of events per 20~GeV bin shown on the vertical axis.  The region \Sleuth\ determines to be most interesting is shown with the (blue) arrow to be the region corresponding to $\sumPt>469$~GeV.  A magnified view of this region in logarithmic scale is shown in the inset, together with the number of events predicted (SM) and the number of events observed ($d$).  Taking into account the trials factor associated with looking at many different regions, \Sleuth\ determines the fraction of hypothetical similar experiments that would produce something in this final state as interesting as the region shown to be $\scriptP=0.0055$ (number at upper right). }
\label{fig:MostInterestingSleuthFinalState}
\end{figure}

\section{Results}
\label{sec:results}

In 927~pb$^{-1}$ of CDF Run II data, the most interesting region is observed in the $b\bar{b}$ final state, shown in Fig.~\ref{fig:MostInterestingSleuthFinalState}.  The fraction of pseudo experiments in this final state alone that would produce a region more interesting than the one observed is $\scriptP=0.0055$.  Taking into account the many final states considered, the fraction of hypothetical similar CDF experiments that would observe a final state with $\scriptP < 0.0055$ is $\tildeScriptP = 0.46$.  

Unfortunately, with $\tildeScriptP = 0.46 \gg 0.001$, \Sleuth\ has not indentified any final state containing a region at large \sumPt\ with a significant excess of data above Standard Model prediction.  

\section{Conclusion}

These proceedings have described \Sleuth, a quasi-model-independent search algorithm designed to identify an excess of data above Standard Model prediction at large summed scalar transverse momentum in any final state.  A specific definition of regions considered allows \Sleuth\ to rigorously calculate the trials factor associated with looking in many different places.  Sensitivity tests performed indicate \Sleuth\ is broadly comparable to targeted searches for specific models adequately satisfying the three assumptions on which it is based.  \Sleuth\ is expected to be significantly more sensitive to unspecified new physics than any particular targeted search, due to \Sleuth's noticeably broader scope.  Application of \Sleuth\ in association with \Vista~\cite{vista} to the first 927~pb$^{-1}$ of CDF II data has unfortunately revealed no indication of new physics.  This analysis represents one of the most encompassing searches so far conducted for new physics at the energy frontier. 

This result does not prove that no new physics is hiding in Tevatron data; merely that this particular global analysis strategy has not yet revealed a discrepancy on which a discovery claim can be based in the 927~pb$^{-1}$ so far analyzed.  A global analysis of more recently collected CDF data is in progress.


%
%

\end{document}